\documentclass[pre,aps,showpacs,12pt,reprint,superscriptaddress,floatfix]{revtex4-1}
\usepackage[mathletters]{ucs}
\usepackage[utf8]{inputenc}
\usepackage{mathtools} % Pour les outils mathématiques avancés
\usepackage{graphicx}
\graphicspath{{/home/praux/Documents/Dossier/These/Publications/thermoelectricity/tex/Thermoelectricité_article1/graphics}}
\usepackage{amsmath}
\usepackage{amsfonts}
\usepackage{amssymb,yfonts,bbm}
\usepackage[bbgreekl]{mathbbol}
\usepackage{mathrsfs}   % \mathscr{...}  lettre cursive majuscule
\usepackage{color}					% Enable  textcolor
\usepackage{bm}      				% Lettre grasses
\usepackage{chemarrow}				% chemical arrows
\usepackage{booktabs}				
\usepackage{array}					
\usepackage[bbgreekl]{mathbbol}  	% to do letter with double bars use:  $\mathbb{ABCdef123}$
\usepackage{slashed}  				% to do letter with diagonal bar
\usepackage[hidelinks]{hyperref}	% create link within the document (to biblio and equations)
\hypersetup{colorlinks=true,citecolor=blue,urlcolor=blue} 
\usepackage{tikz}					% to draw nice graph and sketch
\usepackage{verbatim}
\usepackage{notes2bib}				% to personalize the notes
\bibnotesetup{note-name =,use-sort-key = false}
\DeclareMathAlphabet{\mathpzc}{OT1}{pzc}{m}{it}
\usepackage{svg}        % svg figures
\usepackage{enumitem}

\newtagform{hp}{\, (HP\;}{)}
\newtagform{eg}{\, (EG\;}{)}
\newtagform{hhp}{\, (HHP\;}{)}
\newtagform{chp}{\, (CHP\;}{)}
\newtagform{eg,hhp}{\, (EG,\;HHP\;}{)}

\begin{document}
\title{Thermodynamic Circuits 2: Nonequilibrium conductance matrix for a thermoelectric converter}

\author{Paul Raux}
\affiliation{Université Paris-Saclay, CNRS/IN2P3, IJCLab, 91405 Orsay, France }
\affiliation{Université Paris Cité, CNRS, LIED, F-75013 Paris, France}
\author{Christophe Goupil}
\affiliation{Université Paris Cité, CNRS, LIED, F-75013 Paris, France}
\author{Gatien Verley}
\affiliation{Université Paris-Saclay, CNRS/IN2P3, IJCLab, 91405 Orsay, France }

\date{\today}

\begin{abstract}
In the linear regime, Onsager's response matrix provides the coupling between heat and charge currents crossing a section of thermoelectric materials of infinitesimal thickness. Integrating this response over the finite thickness of a one-dimensional Thermoelectric Converter (TEC) leads to quadratic heat-force characteristics (Joule's law) and linear current-voltage characteristics (Ohm's law). However, these non-linear characteristic equations are not matrix relation anymore. This prevents from determining the currents degree of coupling, albeit its central role for optimizing energy conversion. 
Based on current conservation laws, i.e., the linear dependence between internal physical currents (crossing a section of material) or between external ones (exchanged with the environment), we distinguish two relevant basis of physical and fundamental currents. For those, we define non-equilibrium conductance matrices providing the current-force relations of a TEC in any convenient basis. In doing so, we introduce a degree of coupling between heat and charge currents, in line with the work of Kedem and Caplan but beyond weakly irreversible thermodynamics. This demonstrates by example that non-equilibrium conductance matrices constitute effective models for driven systems, as Onsager response matrices do in the linear regime. 
The sequel papers of this series focus on associating systems modeled in such way.
\end{abstract}
\maketitle
%
%\section{Introduction}
%
The seminal paper~\cite{Onsager1931} of L. Onsager lead to an intense work aiming at discovering the principles of close-to-equilibrium thermodynamics (also called weakly irreversible thermodynamics)~\cite{Schnakenberg1976, Book_Hill1989, Peusner1985, Broeck2015}.
%In his seminal 1931 article~\cite{Onsager1931}, Lars Onsager proposed a close-to-equilibrium thermodynamics framework leading to intense work aiming at discovering the principles of this new field~\cite{Schnakenberg1976, Book_Hill1989, Peusner1985, Broeck2015}. 
Thermoelectric materials have long been regarded as paradigmatic systems in the study of the out-of-equilibrium conversion of heat into electricity. Indeed, in 1948, Callen proposed a concise model for the description of ThermoElectric Converters (TEC) requiring only three parameters called thermoelectric coefficients \cite{Callen1948vol73}. This simple model allowed for an analytical study of the energy and matter transport as well as their coupling. At the mesoscopic scale, these parameters are assumed to remain constant irrespective of thermodynamic forces, an assumption leading to the Constant Property Model (CPM). This modeling approach was extended in 1959 for thermoelectric material of macroscopic thickness by A. Ioffe to incorporate dissipative terms and explicit entropy balances with thermostats~\cite{Ioffe1959}. Nowadays, industrials still use the characteristic equations derived by Ioffe to design efficient TEC, either dithermic engines producing electric power or, in reverse operating mode, to pump heat against thermal gradient at the expense of electric work. 

% Le but de cet article est de définir la matrice de conductance pour un TEC

Before associating TECs made with different (or identical) materials as done in the next paper of this series, we focus on defining the notion of nonequilibrium conductance matrix for a TEC without assuming any underlying microscopic dynamics~\cite{Esposito2015_vol91, Book_Rax2015}. Our goal is to demonstrate that the non-linear characteristics derived by A. Ioffe can be cast in matrix form and represents an effective modeling of a TEC. The choice of conductance matrix amounts to a choice of currents coupling \cite{Vroylandt2018vol2018}, similar to the figure of merit in the linear regime. The corresponding effective description could arise from a microscopic dynamics as well, upon an accurate dynamical parameter tuning. We remark that introducing an underlying dynamics is only required for predicting current fluctuations for instance. % which goes beyond the scope of the present work.  
Hence, modeling with non-equilibrium conductance matrices allows arbitrary degrees of coupling, beyond the binary notion of strong coupling (i.e., proportional currents with same proportionality factor for all forces), while remaining at the level of irreversible thermodynamics. In this framework, according to the chosen thermodynamic conversion, it is relevant to express the conductance matrix for the right set of currents. Changing of currents basis is subtle due to the convective and conductive nature of energy current. Exhibiting a conductance matrix for any basis of currents and thermodynamic forces, with the proof of concept of their effective modeling ability, are our main results. We emphasize that the non-equilibrium conductance matrices obtained in the present work follows from the integration of a linear local model, as shown in appendix~\ref{sec : from local to global}: as such, this second paper does not illustrate yet the method introduced in our first paper that deals with the association of nonlinear devices. In contrast, it features the non-equilibrium conductance as a versatile tool for irreversible thermodynamic modelling.
%This conductance is the matrix analogous to the scalar conductance in stationary electronics. It encapsulates the impedance of a TEC in a single object. 

%Thanks to nonequilibrium conductance matrices (written for the desired pair of fundamental currents), an arbitrary degree of coupling can be defined, going beyond the binary notion of strong coupling. 

%However, this matrix can be used for the associations (serial and parallel) of two TECs as studied in the third paper of this series~\cite{Raux2024vol110}.

The paper is organized as follows:
In section~\ref{sec : non equilibrium conductance matrix}, we delves first into the internal structure of the non-equilibrium conductance matrix. We show that the conductance matrix at the level of fundamental currents (linearly independent) can be read as a sub-block of the conductance matrix at the level of physical currents (linearly dependent currents due to conservation laws). Second, given that energy is transported conductively and convectively, we define and relate external and internal physical currents. Third, we relates those currents to the fundamental ones. This section illustrates on a TEC the selection of fundamental currents and forces developed in the first paper of this series~\cite{Raux2024vol110}, but extends it also to conservation laws with non integer ($0$ or $1$) values. 
%Direct extraction of the nonequilibrium conductance matrix for any suitable choice of fundamental currents is easily achieved based on this section's results. 
Starting directly from Ioffe's equations, section \ref{CondMatrix} provides a conductance matrix for fundamental currents and the corresponding one for physical currents, either internal or external. 

In Appendix \ref{sec : structure of the selection matrix}, we justify the specific form of the so-called selection matrix relating physical and fundamental currents in the first section. Appendix \ref{sec : non unicity of the conductance matrices} illustrates the fact that conductance matrices are non-unique, with the extra degrees of freedom, function of thermodynamic forces, allowing for various coupling between the currents. Finally, Appendix \ref{sec : from local to global} provides an original derivation of the current-force characteristics of a TEC (Ioffe's equations) starting from the local flux-force relation. It showcases how a global-scale framework in thermoelectricity unveils space invariants quantities such as the $F$ functions which characterizes the free fraction of energy in the CPM. 
%Appendix section~\ref{optima} is finally devoted to the optimization of the TEG across various operating regimes (from electric power generation to heat pump systems) in connection with the many representations of a TEC's characteristic equations in different basis of fundamental currents.
%
\section{Levels of description}
\label{sec : non equilibrium conductance matrix}
%

%\textcolor{blue}{We provide now a short description of section III C at the beginning of section III.}

In the first paper of this series \cite{Raux2024vol110}, we used different levels of description of thermodynamic devices in view of connecting them. The fundamental level provides a non redundant basis of currents and forces~\cite{Polettini2016_vol94}. The level of physical currents and forces, although linearly related, is also of interest: First, it determines all the exchanges with the system's environment; second, so as to change of basis of fundamental currents. 

In section \ref{SelectionMatrix}, we exhibit a useful form of the selection matrix that relates physical and fundamental currents. This leads us to a specific structure %(that will appear later) 
of the nonequilibrium conductance matrix at the level of physical currents. In section \ref{subsec : non independent currents for the TEC}, we identify two sets of currents for the description of a TEC: the internal currents crossing the thermoelectric device and the external currents exchanged with the environment. In section~\ref{fundacurrentforce}, we provide two examples of selection matrices allowing to switch from internal (respectively external) physical currents and forces to internal (respectively external) fundamental currents and forces. 
\subsection{Structure of the conductance matrices}
\label{SelectionMatrix}

%For self consistency, we remind below some results of Ref.~\cite{Raux2024vol110,Polettini2016_vol94,Vroylandt2018vol2018} describing the selection of linearly independent currents (so-called fundamental) from a redundant set of currents (so-called physical) using conservation laws and an associated selection matrix. We apply these results on both the internal and the external current sets.  

Following the notation of Ref.~\cite{Raux2024vol110}, we denote $\bm i$ the vector of physical currents and $\bm \ell$ the matrix whose lines are the $|\mathscr{L}|$ linearly independent conservation laws relating the $|\mathscr{P}|$ currents in $\bm i$. This writes in matrix form
\begin{equation}
\bm \ell \bm i=\bm 0.
\label{eq : conservation laws}
\end{equation}
Note that if $\bm \ell$ is not full row rank, it should be reduced to a full row rank matrix by removing appropriated raws. The rank-nullity theorem states that the dimension of the kernel of $\bm \ell$ is $\mathrm{dim(\mathrm{ker}(\bm \ell))}=|\mathscr{P}|-|\mathscr{L}| = |\mathscr{I}|$ that is the number of linearly independent currents (i.e., fundamental currents). Eq.~\eqref{eq : conservation laws} thus means that
\begin{equation}
\bm i = \bm S \bm I \quad \text{ with } \quad  \bm \ell \bm S=\bm 0,
\label{eq : selection of independent currents}
\end{equation}
where we have introduced the selection matrix $\bm S$ whose columns is a possible basis of $\mathrm{ker}(\bm \ell)$ and the fundamental currents vector $\bm I$ with $|\mathscr{I}|$ components. 
We show in Appendix \ref{sec : structure of the selection matrix} that the selection matrix can be written as:
\begin{equation}
\bm S=
\begin{bmatrix}
\mathbb{1}_{|\mathscr{I}|}\\
\bm T
\end{bmatrix},
\label{eq : structure of S}
\end{equation}
with $\mathbb{1}_{|\mathscr{I}|}$ the identity matrix of dimension $|\mathscr{I}|$ and $\bm T$ a matrix that only depends on $\bm \ell$. The above form of selection matrix requires that the order of the components of $\bm i$ is chosen such that
\begin{equation}
\bm i =
\begin{pmatrix}
\bm I\\
\bm i_\mathrm{d}
\end{pmatrix}.
\label{eq : splitting i}
\end{equation}
The vector $\bm i_\mathrm{d}$ of the last $|\mathscr{L}|$ components of $\bm i$ are linear functions of $\bm I$. Since the Entropy Production Rate (EPR) is independent of the level of description, we have
\begin{equation}
\sigma=\bm a^T \bm i=\bm A^T\bm I
\label{eq : EPR dependent = independent}
\end{equation}
$\sigma$ then writes as a function of the conjugated physical force $\bm a$ conjugated to current $\bm i$, or similarly of the fundamental force $\bm A$ and current $\bm I$.
Using Eq.\eqref{eq : selection of independent currents} in Eq.~\eqref{eq : EPR dependent = independent} yields
\begin{equation}
\bm A^T=\bm a^T \bm S.
\label{eq : relation fundamental forces physical forces}
\end{equation}
Then, for a current-force relations reading
\begin{align}
\bm i&=\bm g \bm a, 
\label{eq : physical flux force relation} &&\text{(physical level)} \\
\bm I &= \bm G \bm A,
\label{eq : fundamental current-force relation} && \text{(fundamental level)},
\end{align}
the nonequilibrium conductance matrices $\bm g$ and $\bm G$, at respectively the physical and fundamental levels, are finally linked by
\begin{equation}
\bm g = \bm S \bm G \bm S^T = 
\begin{bmatrix}
\bm G & \bm G \bm T^T\\
\bm T\bm G & \bm T \bm G \bm T^T
\end{bmatrix}.
\label{eq : physical conductance matrix vs fundamental conductance matrix}
\end{equation}
This follows from inserting Eqs.~\eqref{eq : relation fundamental forces physical forces} and \eqref{eq : fundamental current-force relation} in Eq.~\eqref{eq : selection of independent currents}. Therefore, to identify the fundamental conductance matrix knowing the physical conductance matrix in a conveniently chosen basis, it suffices to read the upper left diagonal sub-matrix of dimension $|\mathscr{I}|$. 
\subsection{Physical currents and forces: external, internal}
\label{subsec : non independent currents for the TEC}
\begin{figure}[t]
\centering
\includegraphics[width=0.9\columnwidth]{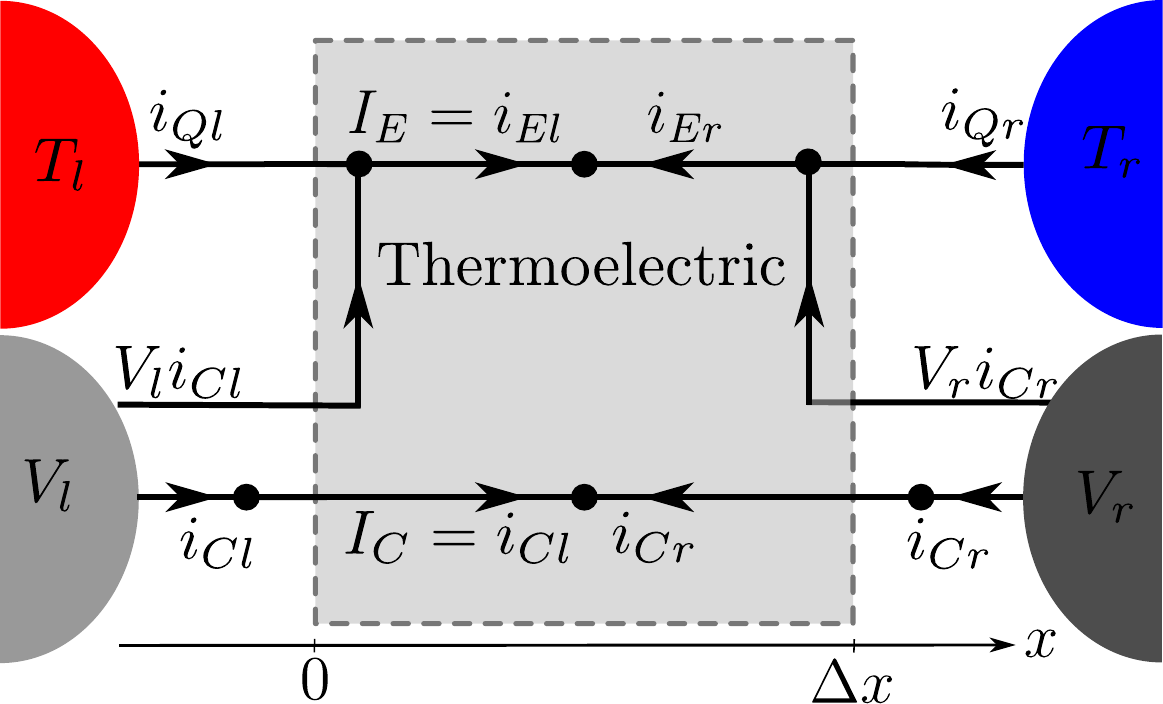}
\caption{The TEC is connected to two thermostats at temperature $T_l$ and $T_r$ with $\Delta T = T_{r}-T_{l} < 0$ and to two metallic leads %chemostats at chemical potential 
at electric potentials $V_l$ and $V_r$ with $\Delta V=V_r-V_l>0$. At each black dots, the Kirchoff current law can be applied. Thus at the interface with the reservoir, there exists four conservation laws. Inside the TEC, two conservation laws are available.}
\label{fig : schema lois de conservation}
\end{figure}
Our sign convention for currents is given on Fig.~\ref{fig : schema lois de conservation}: physical currents are positive when received by the thermoelectric material in the interval $[0,\Delta x]$. Currents entering from the reservoir on the $\chi=l,r$ side (either left or right) are: the (electric) charge current $i_{C\chi}$, the heat current $i_{Q\chi}$ and the energy current $i_{E\chi}$. The electric and energy currents incoming from the left are opposite to those incoming from the right:
\begin{align}
i_{El} =&\,  I_E = - i_{Er} \label{energyconservation}\\
i_{Cl} =& \, \;I_C\; = - i_{Cr}.
\end{align}
Then, the heat current incoming from the $\chi$ side writes
\begin{eqnarray}
	i_{Q\chi} & =& i_{E\chi}-V_{\chi} i_{C\chi},
\end{eqnarray}
Energy conservation of Eq.~\eqref{energyconservation} implies that the work current received by the TEC is
\begin{equation}
	i_{W} \equiv -i_{Ql}-i_{Qr} = -I_C \Delta V \label{workdefinition}
\end{equation}
The conservation laws of physical currents introduced above can be considered at the interface between the system and its environment (grey dashed line in Fig.~\ref{fig : schema lois de conservation}) or across a plane of constant $x\in [0,\Delta x]$. To distinguish them, we call the former \emph{external} physical currents denoted 
\begin{equation}
\bm i_\mathrm{e}^T\equiv
\begin{pmatrix}
i_{Ql} & V_li_{Cl} & i_{Cl} & i_{Qr} & V_r i_{Cr} & i_{Cr} 
\end{pmatrix}, \label{externalcurrents}
\end{equation}
and the latter the \emph{internal} physical currents denoted
\begin{equation}
\bm i_\mathrm{i}^T=
\begin{pmatrix}
i_{El} & i_{Cl} & i_{Er} & i_{Cr} 
\end{pmatrix}. \label{internalcurrents}
\end{equation}
Accordingly, the subscripts $\mathrm{e}$ for external and $\mathrm{i}$ for internal will be used on conjugated and fundamental variables as well. In practice, external currents are useful to make balance with the surrounding leading for instance to the received work above. The internal currents are convenient for the serial association of TECs that we will study in the next paper of this series. In any case, there is no conceptual difference between internal and external currents. For instance, one can define electric current, heat current and electric power across any transverse plane of the thermoelectric material.
The conservation laws for internal currents (electric charge and energy conservation) write
\begin{equation}
\bm\ell^i\bm i_\mathrm{i}=\bm 0, \; \text{with } \bm \ell^i=
\begin{bmatrix}
1 & 0 & 1 & 0\\
0 & 1 & 0 & 1
\end{bmatrix},
\label{eq : conservation laws internal currents}
\end{equation}
and those for external currents writes
\begin{equation}
\bm \ell^e \bm i_\mathrm{e}=\bm 0, \; \text{with } \bm \ell^e =
\begin{bmatrix}
1 & 1 & 0 & 1 & 1 & 0\\
0 & 0 & 1 & 0 & 0 & 1\\
0 & 1 & -V_l & 0 & 0 & 0\\
0 & 0 & 0 & 0 & 1 & -Vr
\end{bmatrix}.
\label{eq : conservation laws external currents}
\end{equation}
Electric charge and energy conservation are complemented by two additional conservation laws that take into account the proportionality between the ``electro-chemical'' works $V_\chi i_\chi$ and the electric currents $i_\chi$ on each $\chi=l,r$ side. Finally, we switch between internal and external currents using
\begin{equation}
\bm i_\mathrm{i}=
\bm  P
\bm i_\mathrm{e},\quad \bm i_\mathrm{e}=\bm M \bm i_\mathrm{i},
\label{eq : relation internal physical currents vs external physical currents}
\end{equation}
where 
\begin{equation}
\bm P=
\begin{bmatrix}
1 & 1 & 0 & 0 & 0 & 0\\
0 & 0 & 1 & 0 & 0 & 0 \\
0 & 0 & 0 & 1 & 1 & 0\\
0 & 0 & 0 & 0 & 0 & 1
\end{bmatrix},\quad 
\bm M=
\begin{bmatrix}
1 & -V_l & 0 & 0 \\
0 & V_l & 0 & 0\\
0 & 1 & 0 & 0\\
0 & 0 & 1 &-Vr\\
0 & 0 & 0 & Vr\\
0 & 0 & 0 & 1
\end{bmatrix}.
\end{equation}
We notice that $\bm P \bm M = \mathbb{1}_{4}$, but $\bm M \bm P \neq \mathbb{1}_{6} $ even though $\bm i_\mathrm{e}=\bm M  \bm P \bm i_\mathrm{e} $ is verified for the vector of external currents of Eq.~\eqref{externalcurrents}. We also emphasize that, although $\bm P$ has linearly independent lines and $\bm M$ has linearly independent columns, one should not use their respectively right and left Moore-Penrose pseudo-inverse. This is clear for matrix $\bm M$ that involves the electric potential which creates problems of physical dimensions when using pseudo-inverses. More importantly, and contrarily to what would come from using pseudo-inverse, matrices $\bm P$ and $\bm M$ provides the appropriated forces conjugated to physical currents in the EPR. Indeed, this rate is independent of the set of variables used to express it
\begin{equation}
\sigma=\bm a_\mathrm{e}^T \bm i_\mathrm{e}=\bm a_\mathrm{i}^T \bm i_\mathrm{i}.
\label{eq : EPR external currents}
\end{equation}
In the stationary state, the EPR is 
\begin{equation}
\sigma=-\sum_{\chi=l,r}\frac{i_{Q\chi}}{T_{\chi}}=-\sum_{\chi=l,r} \frac{i_{E\chi}-V_\chi i_{C\chi}}{T_\chi}.
\end{equation}
This leads to the physical forces conjugated to internal currents
\begin{equation}
\bm a^T_\mathrm{i}=
\begin{pmatrix}
-\frac{1}{T_l} & \frac{V_l}{T_l} & -\frac{1}{T_r} & \frac{V_r}{T_r}
\end{pmatrix}.
\label{eq : physical forces conjugated to the internal currents}
\end{equation}
Using the first relation of Eq.~\eqref{eq : relation internal physical currents vs external physical currents} in Eq.~\eqref{eq : EPR external currents} leads to the physical force conjugated to external currents
\begin{equation}
\bm a_\mathrm{e}^T=\bm a_\mathrm{i}^T\bm P,
\end{equation}
reading explicitly
\begin{equation}
\bm a_\mathrm{e}^T=
\begin{pmatrix}
-\frac{1}{T_l} & -\frac{1}{T_l} & \frac{V_l}{T_l} & -\frac{1}{T_r} & -\frac{1}{T_r} & \frac{V_r}{T_r}
\end{pmatrix}. \label{externalforces}
\end{equation}
Now using the second relation of Eq.~\eqref{eq : relation internal physical currents vs external physical currents} in Eq.~\eqref{eq : EPR external currents}, we obtain a second relation for the forces:
\begin{equation}
\bm a_\mathrm{i}^T=\bm a_\mathrm{e}^T \bm M \label{internalexternalforces}
\end{equation}
which recovers Eq.~\eqref{eq : physical forces conjugated to the internal currents} ensuring the consistency between the description of the TEC using internal or external currents.
\subsection{Fundamental currents and forces: external, internal \label{fundacurrentforce}}
In the previous section, we have identified the physical currents and conjugated forces (internal and external). In this section, we provide some possible fundamental currents and forces by choosing selection matrices in the kernel of the matrix of conservation laws. We could proceed directly by inspection of the EPR, but we aim here at illustrating the general method that is convenient in more involved situations.

Let's start by choosing some fundamental currents and forces among external physical currents. There are $|\mathscr{P}_\mathrm{e}|=6$ linearly dependent external currents and $|\mathscr{L}_\mathrm{e}|=4$ associated conservation laws. We can thus choose $|\mathscr{I}_\mathrm{e}| = |\mathscr{P}_\mathrm{e}|-|\mathscr{L}_\mathrm{e}|=2$ linearly independent currents by removing $|\mathscr{L}_\mathrm{e}|=4$ currents from $\bm i_\mathrm{e}$ (one per conservation law). 
For instance, we can choose for fundamental currents
\begin{equation}
\bm I_\mathrm{e}=
\begin{pmatrix}
i_{Ql}\\
V_l i_{Cl}
\end{pmatrix}, \label{ChoiceFundExt}
\end{equation}
The selection matrix associated to this choice is:
\begin{equation}
\bm S_\mathrm{e}=
\begin{bmatrix}
\mathbb{1}_2\\
\bm T_\mathrm{e}
\end{bmatrix}, \, \text{ with } \bm T_\mathrm{e}=
\begin{bmatrix}
0 & {1}/{V_l}\\
-1 & {\Delta V}/{V_l}\\
0 & -{V_r}/{V_l}\\
0 & -1/V_{l}
\end{bmatrix}. \label{externalselectionmatrix}
\end{equation}
One can check that $\bm \ell_\mathrm{e}\bm S_\mathrm{e}=\bm 0$. The thermodynamic forces conjugated to the fundamental currents of Eq.~\eqref{ChoiceFundExt} follows from Eq.~\eqref{eq : relation fundamental forces physical forces}
\begin{equation}
\bm A_\mathrm{e}=
\begin{pmatrix}
\frac{1}{T_r}-\frac{1}{T_l}\\
-\frac{1}{T_r}\frac{\Delta V}{V_l}
\end{pmatrix}.
\end{equation}

Let's continue with choosing some fundamental currents and forces among internal physical currents. There are $|\mathscr{P}_\mathrm{i}|=4$ linearly dependent internal currents and $|\mathscr{L}_\mathrm{i}|=2$ associated conservation laws. As above, we choose $|\mathscr{I}_\mathrm{i}|  = |\mathscr{P}_\mathrm{i}|-|\mathscr{L}_\mathrm{i}|=2$ linearly independent currents by removing $|\mathscr{L}_\mathrm{i}|=2$ currents from $\bm i_\mathrm{i}$ (one per conservation law). For instance, we can choose for fundamental currents
\begin{equation}
\bm I_\mathrm{i}=
\begin{pmatrix}
i_{El}\\
i_{Cl}
\end{pmatrix} = \begin{pmatrix}
I_{E}\\
I_C
\end{pmatrix}. \label{ChoiceFundInt}
\end{equation}
The selection matrix associated to this choice is:
\begin{equation}
\bm S_\mathrm{i}=
\begin{bmatrix}
\mathbb{1}_2\\
-\mathbb{1}_2
\end{bmatrix} \label{internalselectionmatrix}
\end{equation}
We notice that $\bm S_\mathrm{i}$ has the structure given in Eq.~\eqref{eq : structure of S} with $\bm T_\mathrm{i}=-\mathbb{1}_2$. Here again, one can check that $\bm \ell_\mathrm{i}\bm S_\mathrm{i}=\bm 0$. The thermodynamic forces conjugated to the fundamental currents of Eq.~\eqref{ChoiceFundInt} follows 
\begin{equation}
\bm A_\mathrm{i} =
\begin{pmatrix}
\frac{1}{T_r}-\frac{1}{T_l}\\
\frac{V_l}{T_l}-\frac{V_r}{T_r}
\end{pmatrix} \equiv \begin{pmatrix}
A_{E}\\
A_{C}
\end{pmatrix} ,
\end{equation}
where we have introduced for later convenience the fundamental force $A_{E}$ (respectively $A_{C}$) conjugated to energy (respectively electric) current. By construction, the EPR of the TEC is
\begin{equation}
    \sigma = \bm a_\mathrm{e}^T \bm i_\mathrm{e}=\bm A_\mathrm{e}^T \bm I_\mathrm{e}= \bm a_\mathrm{i}^T \bm i_\mathrm{i}=\bm A_\mathrm{i}^T  \bm I_\mathrm{i}.
    \label{eq : entropy production}
\end{equation}
For systems, with a small number of currents and forces, one can use the conservation laws directly in this EPR to produce a linear combination of fundamental currents only, with each linear coefficient being the conjugated force.

\section{Conductance matrices for physical and fundamental variables}
\label{CondMatrix}
In this section, starting from the Ioffe's current-force characteristics for a CPM, we introduce the conductance matrix of  a TEC at fundamental level. We then go on with the derivation of the conductance matrix at all level of description. In particular, we show how to derive the conductance matrix at fundamental scale coupling heat and electric current. 

%We consider a TEC made of homogenous material, assume a CPM and a effective one dimensional description. 
%The thermoelectric coefficients are $\kappa_J$ the 
According to Ref.~\cite{Apertet2012}, a TEC of finite thickness $\Delta x$ and section area $\mathcal{A}$ admits the following current-force characteristics 
\begin{align}
I_E&=-K\Delta T + F I_C,\label{eq main : global energy current}\\
I_C&=-\frac{\alpha \Delta T + \Delta V}{R},\label{eq main : global matter current}
\end{align}
where the free fraction of transported energy is $F=\alpha \overline{T} + \overline{V}$, with the mean temperature $\overline{T} = (T_l+T_r)/2$ and electric potential $\overline{V} = (V_l+V_r)/2$, and the Seebeck coefficient $\alpha$. We denote $K=\kappa_J\mathcal{A}/\Delta x $ the thermal conductivity under zero electric current and $R=\Delta x/(\sigma_T\mathcal{A})$ the isothermal electric resistance (with $\kappa_J$ and $\sigma_T$ the respective local conductances). We refer to Appendix~\ref{sec : from local to global} for an original derivation of these characteristic equations. 
Eqs.~(\ref{eq main : global energy current}--\ref{eq main : global matter current}) lead to a current-force characteristic in matrix form
\begin{equation}
\bm I_\mathrm{i} =\begin{pmatrix}
I_{E}\\
I_C
\end{pmatrix} = 
-\frac{1}{R}
\begin{bmatrix}
 \alpha F + KR  &   F \\
 \alpha & 1
\end{bmatrix}
\begin{pmatrix}
\Delta T \\
\Delta V
\end{pmatrix}.
\label{eq : Onsager delta T delta mu}
\end{equation}
that involves a non-symmetric matrix and non-conjugated currents and forces. We recover a symmetric matrix by switching to conjugated forces via 
\begin{equation} 
 \begin{pmatrix}
\Delta T \\
\Delta V
\end{pmatrix} =-\frac{T_rT_l}{\bar T} 
\begin{bmatrix}
\bar{T} & 0 \\
\bar{V}  & {1}
\end{bmatrix}
\bm A_\mathrm{i}
\label{eq : force change of basis}
\end{equation}
yielding the current-force relation 
\begin{equation}
\bm I_\mathrm{i}
=\bm G_\mathrm{i}
\bm A_\mathrm{i}
\label{eq : fundamental current-force relation for internal currents}
\end{equation}
with
\begin{equation}
\bm G_\mathrm{i}
=\frac{T_lT_r}{R\bar{T}}
\begin{bmatrix}
F^2 + KR \bar{T} & F \\
F & 1
\end{bmatrix}.
\label{eq : global flux force relation conjugated}
\end{equation}
This conductance matrix is symmetric and positive definite, as required for a TEC arbitrarily far from equilibrium. %Since $F$ depends on $I_C$, it can be expressed in terms of $A_E$ and $A_{C}$, and so does $\bar T$. 
It is also force dependent, here via the $\bar T$ and $\bar V $ appearing in $F$.
Hence, the conductance matrix $\bm G$ provides a non-linear current-force characteristic. 

In Appendix \ref{sec : non unicity of the conductance matrices}, we show that the above conductance matrix is in principle not unique. For given mean currents and conjugated forces, making a choice of conductance matrix amounts to constrain the currents covariance~\cite{Vroylandt2018vol2018}. 
However, the conductance of Eq.~\eqref{eq : global flux force relation conjugated} is the simplest: any other form involves the supplementary definition of a bivariate function of fundamental forces. The introduction of the above nonequilibrium conductance matrices for a TEC is our main result. It is based on the mean currents associated to the constant property model of thermoelectric \cite{Goupil2011_vol13}. Our method, applied here to thermoelectricity, provides a general way of modeling thermodynamic systems without relying on a microscopic model able to predict currents fluctuations. 
%The choice of a nonequilibrium conductance matrix precisely proceeds of a choice of quadratic fluctuations. 
%

We end this section by providing the conductance matrix at the physical level (external and internal) from the one at the fundamental level. First, we apply Eq.~\eqref{eq : physical conductance matrix vs fundamental conductance matrix} using the selection matrix of Eq.~\eqref{internalselectionmatrix} leading to the conductance for internal physical currents
\begin{equation}
\bm g_\mathrm{i}=\bm S_\mathrm{i}\bm G_\mathrm{i}\bm S_\mathrm{i}^T=
\begin{bmatrix}
\bm G_\mathrm{i} & -\bm G_\mathrm{i}\\
-\bm G_\mathrm{i} & \bm G_\mathrm{i}
\end{bmatrix}, \, \text{ for } \bm i_\mathrm{i} = g_\mathrm{i} \bm a_\mathrm{i}.
\end{equation}
As expected this matrix is symmetric, although it is now semi-definite positive only.
%, given the linearly dependent columns and rows. 
Second, we use Eqs.~\eqref{eq : relation internal physical currents vs external physical currents} and \eqref{internalexternalforces} to write
\begin{equation}
	\bm i_\mathrm{e} = \bm M \bm i_\mathrm{i} = \bm M \bm g_\mathrm{i} \bm a_\mathrm{i} = \bm M \bm g_\mathrm{i} \bm M^{T} \bm a_\mathrm{e}%, \;\, \Rightarrow \;\, .
\end{equation}
from which we read the conductance $\bm g_\mathrm{e} = \bm M \bm g_\mathrm{i} \bm M^T$ for external physical currents. We give this conductance matrix in Table \ref{eq: conductance matrix physical external} since it can be used conveniently to extract the conductance matrix at the fundamental level for any choice of fundamental currents.
\begin{table*}
\begin{equation}
\bm g_\mathrm{e} = \frac{T_lT_r}{R\bar{T}}
\begin{bmatrix}
 (F-V_l)^2+ K R\bar{T} & V_l (F-V_l) & F-V_l & -\left((F-V_l) (F-V_r)+KR\bar{T}\right)  &V_r (V_l-F) & V_l-F\\
V_l (F-V_l) & V_l^2 & V_l & V_l (V_r-F) & -V_l V_r & -V_l \\
 F-V_l & V_l & 1 & V_r-F & -V_r & -1\\
 -\left((F-V_l) (F-V_r)+KR\bar{T}\right)  & V_l (V_r-F) & V_r-F &  (F-V_r)^2+ K R\bar{T} & V_r (F-V_r) & F-V_r\\
V_r (V_l-F) & -V_l V_r & -V_r & V_r (F-V_r) & V_r^2 & V_r\\
 V_l-F & -V_l & -1 & F-V_r & V_r & 1
\end{bmatrix}.
\nonumber
\end{equation}
\caption{Conductance matrix for physical currents and forces given in Eqs.~\eqref{externalcurrents} and \eqref{externalforces}. \label{eq: conductance matrix physical external}}
\end{table*}
To do so, one simply select $\bm g$'s components for the lines and columns associated to the chosen currents and forces. We remark that conjugated forces must still be determined using the appropriate selection matrix for chosen currents. For instance, one has the following current-force relation at the fundamental level  
\begin{equation}
\begin{pmatrix}
i_{Ql}\\
i_{Cl}
\end{pmatrix}
=\begin{bmatrix}
	g_{11} & g_{13}\\
	g_{31} & g_{33}
	\end{bmatrix}
\begin{pmatrix}
\frac{1}{T_r}-\frac{1}{T_l}\\
-\frac{\Delta V}{T_r}
\end{pmatrix}.
\end{equation}
We use table \ref{eq: conductance matrix physical external} to read directly the conductance matrix 
\begin{equation}
	\begin{bmatrix}
	g_{11} & g_{13}\\
	g_{31} & g_{33}
	\end{bmatrix}=\frac{T_lT_r}{R\bar{T}}
\begin{bmatrix}
KR\bar{T}+(F-V_l)^2 & F-V_l\\
F-V_l & 1
\end{bmatrix}
\end{equation} 
The degree of coupling $\xi \in [-1,1] $ between the heat and charge currents entering from the left of the TEC, respectively $i_{Ql}$ and $i_{Cl}$, follows 
\begin{equation}
\xi= \frac{g_{13}}{\sqrt{g_{11}g_{33}}} = \frac{F-V_l}{\sqrt{KR\bar{T}+(F-V_l)^2}} .
\end{equation}
It generalizes far from equilibrium the notion of degree of coupling that has a fundamental role in conversion processes \cite{Kedem1965_vol61, Polettini2015_vol114, Vroylandt2018vol2018}.

\section{Conclusion}
Starting from the Ioffe's integrated current-force characteristics of TEC, we summarized the energy/matter current-force characteristics by introducing a conductance matrix relating the currents to their conjugated forces in the entropy production rate. This matrix is not unique since it corresponds to a choice of model for the coupling between matter and energy. Moreover, we gave an algebraic procedure to change the current basis using the conservation laws of the TEC. This procedure shows that two equivalent sets of currents are available for the description of a TEC: the internal currents (energy and matter), and the external currents that can be easily determined by measuring the currents (heat and matter) entering the boundaries of the TEC. For each of these current sets, we determined an associated nonequilibrium conductance matrix. In the third paper of this series, we use these nonequilibrium conductance matrices to study the circuit associations of TEC in line with the first paper of the series. We emphasize that nonequilibrium conductance matrices are force-dependent and thus apply far from equilibrium, even in non-linear cases.

%Finally, we revisited the derivation of the optima of a TEC. We considered two non-trivial regimes accessible to the TEC: the EG and the HP mode. In both cases, a spontaneous process (with positive partial EPR) fuels a non-spontaneous process (with negative partial EPR). We clarified the study of the efficiencies in each regime by highlighting the symmetry between the EG and the HP regimes. Indeed, we have shown that the optimisation problem in each regime can be expressed in terms of the EG efficiency only. The latter presents a maximum in the EG regime and a minimum in the HP regime corresponding to maximum HP efficiency. Moreover, we introduce the cost of energy (COE) for a TEC. The latter presents an optimum for both regimes. In the case of the generator mode, it is a minimum that reflects the presence of energy consumption, even at zero charge-current. We should point out that this signature, illustrated here in the thermoelectric case, is present in all animated systems, whether living \cite{goupil2019thermodynamics} or in the field of active matter \cite{Davis2024}. Overall, this derivation of the optima of the TEC sheds light on the symmetry relating the EG regime with the HP regime.
%

\section*{Acknowledgments}
P.R. acknowledges fruitful discussions with Armand Despons.

\appendix
\section{Structure of the selection matrix}
\label{sec : structure of the selection matrix}
In this appendix, we justify the form of the selection matrix given in Eq.~\eqref{eq : structure of S}. The matrix of conservation law $\bm \ell $ is of dimension $|\mathscr{L}| \times |\mathscr{P}|$ and $\mathrm{rank}(\bm \ell)=|\mathscr{L}| $. It describes the $|\mathscr{L}|$ linear relations between the physical currents leading to $\bm \ell \bm i = 0$, keeping in mind that these relations are independent. Given the rank of $\bm \ell$, we can chose the order of physical currents' components of $\bm i$ such that the $|\mathscr{L}|$ last columns of matrix $\bm \ell $, denoted $\bm \ell_{d}$, are linearly independent. The remaining $|\mathscr{I}|$ first columns of $\bm \ell$ are denoted $\bm \ell_{I}$. Then, we take the $|\mathscr{I}|$ first components of $\bm i$ as our choice of fundamental currents such that
\begin{equation}
\bm i =
\begin{pmatrix}
\bm I\\
\bm i_\mathrm{d}
\end{pmatrix} \quad \text{ and } \quad 
\bm \ell = 
%\overset{\overset{|\mathscr{I}|}{\longleftrightarrow} \overset{|\mathscr{L}|}{\longleftrightarrow} }{}
\begin{pmatrix}
\bm \ell_I & \bm \ell_{d}
\end{pmatrix}.
\label{eq : splitting}
\end{equation}
The last $|\mathscr{L}|$ components of $\bm i$, denoted $\bm i_{d}$, are the dependent currents that can be obtained as linear combination of the components of $\bm I$. Eq.~\eqref{eq : splitting} is a specific choice ensuring two properties: First, the square matrix $\bm \ell_{d}$ has rank equal to its dimension $|\mathscr{L}|$ and hence is invertible. Second, the selection matrix relating physical and fundamental currents by $\bm i= \bm S \bm I$ writes
\begin{equation}
\bm S = \begin{pmatrix}
\mathbbm{1} \\
\bm T
\end{pmatrix}, \label{struc selection matrix}
\end{equation}
where $\mathbbm{1}$ is the identity matrix of dimension $|\mathscr{I}|$ and the matrix $\bm T$ follows from $\bm \ell \bm S = 0$ leading to 
\begin{equation}
\bm\ell_I + \bm\ell_{d} \bm T = 0 \quad \Rightarrow \quad \bm T = - \bm \ell_{d}^{-1} \bm \ell_{I}.
\end{equation}
This justifies the form of selection matrix given in Eq.~\eqref{eq : structure of S}.
\section{Non unicity of the conductance matrices}
\label{sec : non unicity of the conductance matrices}
Exhibiting a conductance matrix from the current-force relation cannot be done uniquely without relying on a microscopic model. We illustrate this fact for our model of TEC, starting from the non equilibrium conductance matrix of Eq.~\eqref{eq : global flux force relation conjugated}. 
Let's assume that there exist a symmetric matrix $\bm G_\mathrm{i}^{'}\neq \bm G_\mathrm{i}$ and such that
\begin{equation}
\bm G_\mathrm{i}^{'} = \frac{T_lT_r}{R\bar{T}}\begin{bmatrix}
a & b \\
b & c
\end{bmatrix},
\end{equation}
and for which the current-force characteristics is preserved: 
\begin{equation}
\bm I_\mathrm{i}=\bm G_\mathrm{i} \bm A_\mathrm{i} = \bm G_\mathrm{i}^{'}\bm A_\mathrm{i},
\label{eq : flux equality}
\end{equation}
or equivalently
\begin{equation}
\begin{bmatrix}
KR\bar{T} + F^2  & F\\
F & 1
\end{bmatrix}
\begin{pmatrix}
A_E\\
A_{C}
\end{pmatrix}=
\begin{bmatrix}
a & b\\
b & c
\end{bmatrix}
\begin{pmatrix}
A_E\\
A_{C}
\end{pmatrix}.
\end{equation}
Each line of this equation can be solved for $b$ as function of $a$ and $c$. Equalizing the results, we obtain the following constraint relating $a$ and $c$
\begin{equation}
\frac{A_E}{A_{C}}\left(KR\bar{T} + F^2 - a \right) = \frac{A_{C}}{A_E}\left( 1-c \right).
\end{equation}
Finally, $\bm G_\mathrm{i}^{'}$ can be expressed as
\begin{equation}
\bm G_\mathrm{i}^{'}=\bm G_\mathrm{i}
+ \frac{T_lT_r}{R\bar{T}}
\begin{bmatrix}
-\left(\frac{A_{C}}{A_E}\right)^2 & \frac{A_{C}}{A_E}\\
\frac{A_{C}}{A_E} & -1
\end{bmatrix}
(1-c),
\end{equation}
showing that the nonequilibrium conductance matrix $\bm G_\mathrm{i}$ is not unique when providing the current-force relation only, with no information on the currents fluctuations. Any value of $c$ in the above equation produces a conductance matrix compatible with this current-force relation. We emphasize that $c$ is in principle a function of $A_E$ and $A_{C}$. Close-to-equilibrium, the non-equilibrium conductance coincides by definition with Onsager's response matrix: the knowledge of the currents covariance close to an equilibrium states allows to define the conductance matrix uniquely.
The fact that a non-equilibrium conductance matrix includes more information on the model than the current-force relation is argued in Ref.~\cite{Vroylandt2018vol2018}. A non-equilibrium conductance matrix can arise (with a unique definition) from a microscopic modeling. Another approach is to consider it as an alternative way of defining a model, with the idea that the additional information given on the model characterizes the quadratic fluctuations of currents in different non-equilibrium stationary states.
\section{Current-force characteristics of the CPM}
%\section{Derivation of Eqs \eqref{eq main : global energy current} and \eqref{eq main : global matter current}}
\label{sec : from local to global}
%In such scenarios, identifying space-invariant quantities for the out-of-equilibrium system facilitates the system's description since it reduces the number of degrees of freedom by taking matter and energy conservation into account. For a given CPM describing a TEC, we show that energy and matter fluxes both derive from potentials. Charge and energy conservation then implies that the gradients of these potentials are two space-invariant quantities. This constitutes our first result. 

%
In this section, we start from the local (and linear) description of a TEC in terms of a flux-force characteristics defined by the Onsager matrix and its associated thermoelectric coefficients. Using the conservation of matter and energy fluxes across a surface $\mathcal{A} = |\bm {\mathcal{A}}|$ of an homogeneous thermoelectric material of finite thickness $\Delta x$, we obtain the global current-force characteristic. As compared to Ref.~\cite{Apertet2013_vol88}, our derivation emphasizes that both charge and energy currents derive from potential functions that we identify. We also shed light on the role of a space independent quantity that we denote $F$. Finally, we rephrase Domenicalli's equation as a necessary condition for the conservation of energy.
\subsection{Local Onsager flux-force relation}
Let's start by physically motivating each term in the linear characteristic equations of (an effectively one dimensional) thermoelectric material of infinitesimal thickness $\Delta x$. In such material, the heat flux $J_Q$ (W.m$^{-2}$) and the electrical flux $J_C$ (C.s$^{-1}$.m$^{-2}$) are coupled by the local flux-force relation:
\begin{equation}
\begin{pmatrix}
J_Q\\
J_C 
\end{pmatrix}= \bm L \begin{pmatrix}
-\frac{dT}{dx}\\
-\frac{dV}{dx} 
\end{pmatrix},
\label{eq : electrical heat flux force relation thermoelectric coefficients}
\end{equation}
where the thermodynamic forces are respectively the temperature gradient and the electric potential. The temperature $T=T(x)$ and the electric potential $V=V(x)$ are assumed to be constant in any transverse plane of the material with constant $x$ value. The response matrix appearing in Eq.~\eqref{eq : electrical heat flux force relation thermoelectric coefficients} is
\begin{equation}
\bm L=\begin{bmatrix}
\alpha^2 \sigma_T T + \kappa_J & \alpha \sigma_T T \\
\alpha \sigma_T & \sigma_T \\
\end{bmatrix}, \label{HeatResponseMatrix}
\end{equation}
%
%The transport of heat and electric charge described by Eq.~\eqref{eq : electrical heat flux force relation thermoelectric coefficients} can be interpreted as follows. 
where $\kappa_J$ is the thermal conductivity under zero electrical current and $\sigma_T$ the isothermal electrical conductivity. Indeed, we recover Fourier's law $J_Q=-\kappa_J\frac{dT}{dx}$ associated to conductive heat flux (respectively Ohm's law $J_C=-\sigma_T \frac{dV}{dx}$) by taking $\alpha \rightarrow 0$ in the first (respectively second) row of Eq.~\eqref{eq : electrical heat flux force relation thermoelectric coefficients}. The Seebeck coefficient $\alpha$ couples heat and charge fluxes and is defined by 
\begin{equation}
	\alpha=-\left. \frac{\frac{dV}{dx}}{\frac{dT}{dx}}\right|_{J_C=0}.
\end{equation}
Dimension analysis accounts for the off-diagonal terms in $\bm L$ that must involve $\alpha$ by definition. The first term $\alpha^{2}\sigma_{T} T$ in the first diagonal component of $\bm L$ arises from convective heat flux. Finally, we remark the matrix $\bm L$ is not symmetric: Onsager's reciprocity relation does not hold because the force vector $(-\frac{dT}{dx},-\frac{dV}{dx})$ and flux vector $(J_Q,J_C)$ are not conjugated thermodynamic variables. Onsager's reciprocity relations hold in a basis of conjugated thermodynamic variables: here the force vector $(\frac{d}{dx}\left(\frac{1}{T}\right),-\frac{d}{dx}\left( \frac{V}{T} \right))$ conjugated to the current vector $(J_Q,J_C)$.

%In section \ref{integration}, we integrate the local flux-force relation over a finite thickness of thermoelectric material. This leads to the non-linear current-force characteristic that we use in section~\ref{sec : non equilibrium conductance matrix} to introduce the nonequilibrium conductance of our TEC. 
In the following, we integrate the local flux-force relation over a finite thickness of thermoelectric material. To prepare for this integration, it is convenient to introduce the conserved fluxes $(J_E,J_C)$ of energy and charge conjugated to the thermodynamic forces $( \frac{d}{dx}\left(\frac{1}{T}\right),-\frac{d}{dx}\left(\frac{V}{T}\right))$. The energy flux $J_{E}$ is defined according to the first law of thermodynamics by
\begin{equation}
J_E=J_Q +V(x)J_C
\label{eq : local first law}
\end{equation}
where $V(x) J_C$ is the flux of electric power crossing the surface $ \bm{\mathcal{A}}$ in plane of constant $x$ with unit normal vector oriented toward growing $x$. From the conservation of energy and charge, we remark that the heat flux $J_{Q} = J_E - V(x)J_C$ is $x$ dependent. This is precisely this difference between the heat fluxes across the surface $ \mathcal{A}$ at $x=0$ and at $x=\Delta x$ that allows thermoelectric power generation. In the new basis of flux $(J_E,J_C)$ and forces $( \frac{d}{dx}\left(\frac{1}{T}\right),-\frac{d}{dx}\left(\frac{V}{T}\right))$, the local flux-force relation becomes
\begin{equation}
\begin{pmatrix}
        J_E \\
        J_C
    \end{pmatrix}=\bm{ \mathcal{L}} \begin{pmatrix}
    \frac{d}{dx}\left(\frac{1}{T}\right) \\
    -\frac{d}{dx}\left(\frac{V}{T}\right)
\end{pmatrix},
\label{eq : energy matter flux force relation thermoelectric coefficients}
\end{equation} 
involving the response matrix
\begin{equation}
     \bm{ \mathcal{L}}=
    \begin{bmatrix}
        \kappa_J T^2 + T\sigma_T\left( \alpha T + V \right)^2 & T\sigma_T \left( \alpha T +V \right)\\
        T\sigma_T\left(\alpha T +V \right) & T\sigma_T
    \end{bmatrix}. \label{EnergyResponseMatrix}
\end{equation}
This matrix follows from first using in Eqs.~(\ref{eq : electrical heat flux force relation thermoelectric coefficients}--\ref{HeatResponseMatrix}) the following change of variables 
\begin{equation}
\begin{pmatrix}
- \frac{dT}{dx} \\
- \frac{dV}{dx}
\end{pmatrix}=
\begin{bmatrix}
0 & T^{2} \\
VT & T 
\end{bmatrix}
\begin{pmatrix}
\frac{d}{dx}\left(\frac{1}{T}\right) \\
-\frac{d}{dx}\left(\frac{V}{T}\right)
\end{pmatrix},
\end{equation}
and second by using Eq.~\eqref{eq : local first law} to switch from the vector of heat and charge fluxes to the vector of energy and charge fluxes.
In the following, we call Eq.~\eqref{eq : energy matter flux force relation thermoelectric coefficients} the local flux-force relation for the conserved fluxes. Since we use variables that are conjugated when considering an entropy balance, the matrix $\mathcal{L}$ is symmetric and verifies Onsager's reciprocity relations. It is also $x$ dependent via the local temperature $T$ and electric potential $V$.
\subsection{Current-force relation from integrated fluxes \label{integration}}
In this section, we switch from the local level, with fluxes crossing an infinitesimal slice of thermoelectric material, to the global level, with currents defined as space integrated fluxes over $\mathcal{A}$ and crossing a finite thickness $\Delta x$ of thermoelectric material. This integration step produces a non-linear current-force relation from a linear force-flux relation. When integrating the local flux-force relation, we use Dirichlet boundary conditions for the TEC:
\begin{align}
	T(0)	&=T_l, 	& V(0) 	& =V_l, \label{leftBC}\\
	T(\Delta x)	&= T_r, 	& V(\Delta x) 	&= V_r. \label{rightBC}
\end{align}
Without loss of generality, we assume $T_l>T_r$. By convention, all flux and currents are algebraic and positive when flowing toward the growing $x$ direction. We start by showing that both matter and energy fluxes derive from a potential. Using this property, the global current-force relation is then obtained by integrating the differential equations for those potentials on the finite thickness of thermoelectric material. We then determine the temperature and electric potential profile in the $x$ direction. 
\subsubsection{Electric current}
\begin{figure}[t]
\centering
\includegraphics[width=\columnwidth]{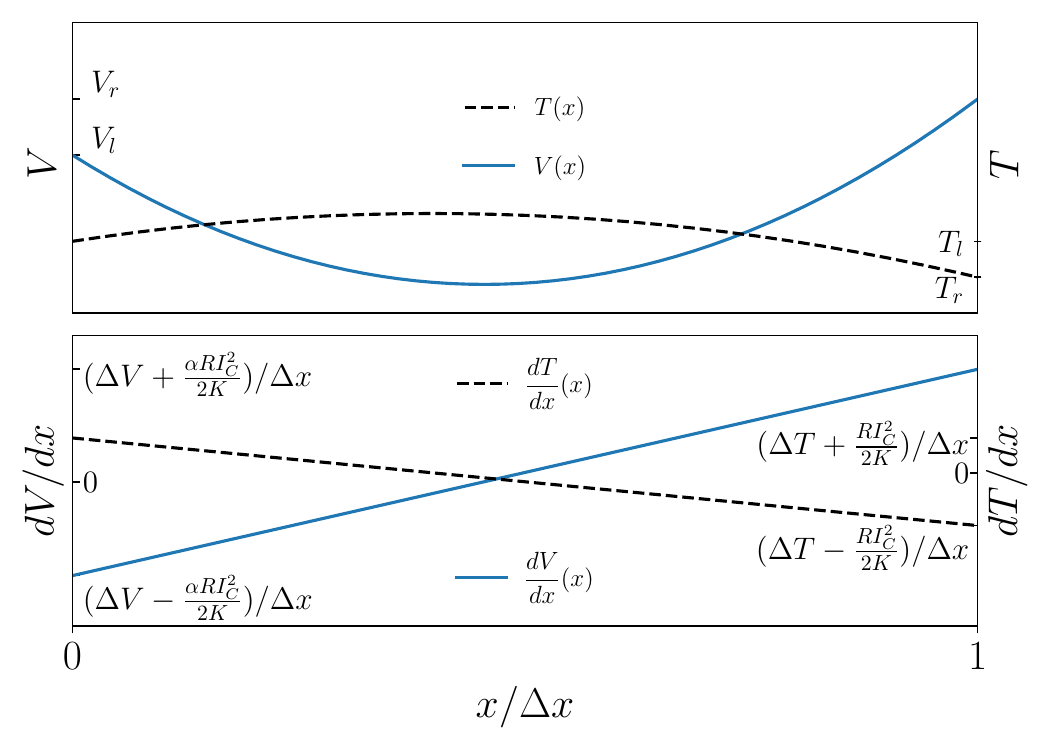}
\caption{ Electric potential and temperature profiles (Top panel) and gradients (Bottom panel) along the $x$ axis of the thermoelectric material under the fixed boundary conditions of Eqs.~(\ref{leftBC}--\ref{rightBC}). The left vertical axis is for electric potential $V$ profile (respectively gradients) in blue solid line. The right vertical axis is for temperature profile (respectively gradients) in black dashed line.
\label{fig : schema gradients}}
\end{figure}

We start by determining the electric current as a function of the temperature difference $\Delta T=T_{r}-T_{l}$ and the tension $\Delta  V=V_r-V_l$ between the right and left planes of the TEC. According to Eqs.~(\ref{eq : electrical heat flux force relation thermoelectric coefficients}--\ref{HeatResponseMatrix}), the electric flux $J_C$ derives from a potential $\varphi(x)$ as
\begin{equation}
    J_C = - \sigma_T \frac{d\varphi}{dx}, \label{eq : electric flux}
\end{equation}
where
\begin{equation}
\varphi=\alpha T + V .
\label{eq : def of phi}
\end{equation}
In the stationary state, this flux is divergence less to ensure charge conservation
\begin{equation}
     \frac{dJ_C}{dx}= - \sigma_T \frac{d^2\varphi}{dx^2}=  0.
    \label{eq : matter conservation}
\end{equation}
From this last equation, we identify the following space independent quantity:
\begin{equation}
    \frac{d}{dx} \left( \alpha T + V \right)=\mathrm{const}.
\end{equation}
In other words, $\varphi$ is an affine function of $x$ with coefficients yet to be determined.
For our effectively one dimensional TEC under the boundary conditions of Eqs.~(\ref{leftBC}--\ref{rightBC}), see Fig.~\ref{fig : schema gradients}, the solution of  Eq.~\eqref{eq : matter conservation} reads
\begin{equation}
\varphi(x) = \left( \varphi_r-\varphi_l \right)\frac{x}{\Delta x} + \varphi_l \label{eq : phi (1)}
\end{equation}
where we have denoted $\varphi_\chi=\alpha T_\chi + V_\chi$ with $\chi=l,r$. Then, the electric current across the oriented surface vector $ \bm{\mathcal{A}} $ follows from the expression of the electric flux \eqref{eq : electric flux} combined with Eq.\eqref{eq : phi (1)} as 
\begin{equation}
I_C\equiv J_C \mathcal{A} =\frac{\varphi_l-\varphi_r}{R}\label{eq : electric current (1)}
\end{equation}
Remarkably, the expression of $I_C$ given in Eq.~\eqref{eq : electric current (1)} takes the form of the Ohm's law generalized to the composite potential $\varphi$ which encapsulates the coupling between energy and charge transport. $\varphi$ can then be expressed in term of the thermoelectric coefficients of the TEC and of the potential of the reservoirs as:
\begin{equation}
\varphi(x)=  \left(\alpha \Delta T + \Delta V\right) \frac{x}{\Delta x} + \alpha T_l + {V_l}.
\label{eq : solution of matter conservation}
\end{equation}
Then, a second expression of the electrical current, given in Eq.~\eqref{eq main : global matter current} of the main text, follows as
\begin{equation}
I_C
%&=-\sigma_T \frac{d}{dx}\left(\alpha(x)T + V(x) \right) A\\
=-\frac{1}{R}\left(\alpha \Delta T + \Delta V\right).
\label{eq : global matter current}
\end{equation}
We emphasize that in generator convention a positive power is generated by the TEC and absorbed by the load when $I_C\Delta V>0$ (for $\Delta V$ and $I_C$ aligned in generator convention and anti-aligned in load convention).

\subsubsection{Energy current}
We continue with the determination of the energy current as a function of $\Delta T$ and $\Delta V$. According to Eqs.~(\ref{eq : electrical heat flux force relation thermoelectric coefficients}--\ref{HeatResponseMatrix}), the heat flux writes 
\begin{equation}
	J_{Q} = \alpha T J_C - \kappa_{J}\frac{dT}{dx} \label{heatflux}
\end{equation} 
that, when combined with Eq.~\eqref{eq : local first law} and the definition of $\varphi$ in Eq.~\eqref{eq : def of phi}, leads to the energy flux 
\begin{equation}
J_E=\varphi J_C - \kappa_J\frac{dT}{dx}.
\label{eq : je en fonction de j et nabla T}
\end{equation}
Let's show that $J_E$ derives from a potential. Indeed, combining Eq.~\eqref{eq : electric flux} and Eq.~\eqref{eq : je en fonction de j et nabla T} yields
\begin{equation}
J_E= - \kappa_J\frac{d\Phi}{dx},
\label{eq : energy flux}
\end{equation}
where 
\begin{equation}
\Phi= T + \frac{\sigma_T}{2\kappa_J}\varphi^2.
\label{eq : def Phi}
\end{equation}
The potential function $\Phi$ depends on $x$ through $T$ and $\varphi$. Energy conservation is thus ensured if and only if 
\begin{equation}
\frac{dJ_E}{dx}=0=-\kappa_J \frac{d^2\Phi}{dx^2}.
\end{equation}
This last equation is readily solved as
\begin{equation}
\Phi(x)=\left( \Phi_r-\Phi_l \right)\frac{x}{\Delta x} + \Phi_l
\label{eq : Phi(x)}
\end{equation}
where we have denoted $\Phi_\chi=T_\chi + \frac{\varphi_\chi^2}{2KR}$ with $\chi=l,r$. The energy current across the oriented surface vector $ \bm{\mathcal{A}}$ then follows from the expression of the energy flux Eq.~\eqref{eq : energy flux} combined with Eq.~\eqref{eq : Phi(x)}
\begin{equation}
I_E\equiv J_E \mathcal{A} =K\left( \Phi_l - \Phi_r \right).
\label{eq : energy current}
\end{equation}
Once again, we remark that the expression of $I_E$ given in Eq.~\eqref{eq : energy current} takes a Fourier's law form with thermal resistance $1/K$ and generalized to the composite potential $\Phi$ which also encapsulates the coupling between charge and energy transport. We can finally express $\Phi$ in term of the thermoelectric coefficients and the potentials of the reservoirs as
\begin{equation}
\Phi(x)=\left(\Delta T - \frac{F}{K} I_C \right) \frac{x}{\Delta x}+ T_l + \frac{(\alpha T_l+V_l)^2}{2KR}
\end{equation}
where we use the $F$ introduced in the main text which reads
\begin{equation}
F=\frac{\varphi_l+\varphi_r}{2}=\alpha\bar{T}+\bar{V},
\end{equation}
with the mean temperature $\bar{T}=(T_l+T_r)/2$ and electric potential $\bar{V}=(V_l+V_r)/2$.
The energy current can thus be rewritten as
\begin{equation}
I_E=-K\Delta T + F I_C
\label{eq : global energy current}
\end{equation}
which is the equation given in Eq.~\eqref{eq main : global energy current} of the main text. Interestingly, $F$ can be expressed in terms of the charge current by using Eq.~\eqref{eq : global matter current} 
\begin{eqnarray}
F &=& \varphi_l-  \frac{RI_C}{2}\label{eq : Fl}, \\
F &=& \varphi_r +  \frac{RI_C}{2}.
\label{eq : Fr}
\end{eqnarray}
As previously mentioned, in the absence of any leakage term, that is when $K=0$, the matter and energy currents are strictly proportional. Hence, the parameter $F$ indeed characterizes the free fraction of transported energy.
Eqs.~(\ref{eq : global matter current},\ref{eq : global energy current}) constitute the integrated current-force characteristics of a TEC where the forces are the temperature and electric potential differences. %Section III summarizes these results into the so-called non-equilibrium conductance matrix. 
\subsection{Temperature and electric potential profile}
We now derive the temperature and the electric potential profile along the \textit{x} direction of the TEC. The temperature is the solution of Domenicalli's equation~\cite{Apertet2013_vol88}. 
Within our formalism, Domenicalli's equation is in fact equivalent to ensuring the conservation of the total energy flux $J_E$ in Eq.~\eqref{eq : je en fonction de j et nabla T}. Indeed, the total energy flux is conserved if it is divergence less $\frac{dJ_E}{dx}=0$ which using Eq.~\eqref{eq : je en fonction de j et nabla T} reads
\begin{equation}
 0=\frac{d}{dx}\left(\alpha T+V\right) J_C + \left(\alpha T+V \right)\frac{dJ_C}{dx} - \kappa_J\frac{d^2T}{dx^2}
\end{equation}
Then inserting the charge conservation Eq.~\eqref{eq : matter conservation} in this last equation and solving for $\frac{d^2T}{dx^2}$ yields Domenicalli's equation as
\begin{equation}
\frac{d^2T}{dx^2} = - \frac{ J_C^2}{\kappa_J\sigma_T}.
\label{eq : Domenicalli equation}
\end{equation}
This equation combined with Eq.~\eqref{eq : matter conservation} gives a similar second order differential equation for the electric potential instead of the temperature
\begin{equation}
\frac{d^2V}{dx^2}= \frac{\alpha J_C^2}{\kappa_J\sigma_T}.
\end{equation}
For our effectively one dimensional TEC under the boundary conditions of Eqs.~(\ref{leftBC}--\ref{rightBC}), the temperature and electric potential read
\begin{eqnarray}
T &=& T_l +  \frac{x}{\Delta x}\left[ \Delta T-\frac{RI_C^2}{2K}\left( \frac{x}{\Delta x}-1\right) \right] \label{potentialprofile}, \\
V &=& V_l +  \frac{x}{\Delta x}\left[ \Delta V+\frac{\alpha RI_C^2}{2K}\left( \frac{x}{\Delta x}-1\right) \right] \label{temperatureprofile}.
\end{eqnarray}
The couple of Eqs.~(\ref{potentialprofile}--\ref{temperatureprofile}) is in agreement with Eq.~\eqref{eq : solution of matter conservation} that led us to the electric current in the previous section. 
From the temperature gradient
\begin{equation}
\frac{dT}{dx} =\frac{1}{\Delta x} \left[\Delta T - \frac{RI_C^2}{2K}\left( \frac{2x}{\Delta x} - 1 \right) \right] %\frac{ \bm{\mathcal{A}}}{\vert  \bm{\mathcal{A}} \vert},
\label{eq : nabla T}
\end{equation}
one can, interestingly revisit the determination of $I_E$ by multiplying  Eq.~\eqref{eq : je en fonction de j et nabla T} and inserting Eq.~\eqref{eq : nabla T} such that
\begin{equation}
I_E=-K\Delta T+ \left(\alpha T +V +\frac{RI_C}{2}\left( \frac{2x}{\Delta x} - 1 \right)\right)I_C .
\label{eq : integrated energy flux}
\end{equation}
The energy conservation across any plane with constant $x$, imposes that the energy current is $x$ independent, meaning that we have exhibited a new expression of $F$ which is a non trivial $x$ invariant quantity within the thermoelectric material. It can be expressed in term of $\varphi$ as 
\begin{equation}
	F(x)\equiv \varphi(x) + \frac{RI_C}{2}\left( \frac{2x}{\Delta x} - 1 \right)=\mathrm{const}.
\end{equation}
We can thus interpret the expressions of $F$ found in Eqs.\eqref{eq : Fl} and \eqref{eq : Fr} as boundary values of the invariant $F(x)$ respectively on the left and on the right sides.
\bibliography{cited_ref}

%merlin.mbs apsrev4-1.bst 2010-07-25 4.21a (PWD, AO, DPC) hacked
%Control: key (0)
%Control: author (8) initials jnrlst
%Control: editor formatted (1) identically to author
%Control: production of article title (-1) disabled
%Control: page (0) single
%Control: year (1) truncated
%Control: production of eprint (0) enabled
\begin{thebibliography}{17}%
\makeatletter
\providecommand \@ifxundefined [1]{%
 \@ifx{#1\undefined}
}%
\providecommand \@ifnum [1]{%
 \ifnum #1\expandafter \@firstoftwo
 \else \expandafter \@secondoftwo
 \fi
}%
\providecommand \@ifx [1]{%
 \ifx #1\expandafter \@firstoftwo
 \else \expandafter \@secondoftwo
 \fi
}%
\providecommand \natexlab [1]{#1}%
\providecommand \enquote  [1]{``#1''}%
\providecommand \bibnamefont  [1]{#1}%
\providecommand \bibfnamefont [1]{#1}%
\providecommand \citenamefont [1]{#1}%
\providecommand \href@noop [0]{\@secondoftwo}%
\providecommand \href [0]{\begingroup \@sanitize@url \@href}%
\providecommand \@href[1]{\@@startlink{#1}\@@href}%
\providecommand \@@href[1]{\endgroup#1\@@endlink}%
\providecommand \@sanitize@url [0]{\catcode `\\12\catcode `\$12\catcode
  `\&12\catcode `\#12\catcode `\^12\catcode `\_12\catcode `\%12\relax}%
\providecommand \@@startlink[1]{}%
\providecommand \@@endlink[0]{}%
\providecommand \url  [0]{\begingroup\@sanitize@url \@url }%
\providecommand \@url [1]{\endgroup\@href {#1}{\urlprefix }}%
\providecommand \urlprefix  [0]{URL }%
\providecommand \Eprint [0]{\href }%
\providecommand \doibase [0]{http://dx.doi.org/}%
\providecommand \selectlanguage [0]{\@gobble}%
\providecommand \bibinfo  [0]{\@secondoftwo}%
\providecommand \bibfield  [0]{\@secondoftwo}%
\providecommand \translation [1]{[#1]}%
\providecommand \BibitemOpen [0]{}%
\providecommand \bibitemStop [0]{}%
\providecommand \bibitemNoStop [0]{.\EOS\space}%
\providecommand \EOS [0]{\spacefactor3000\relax}%
\providecommand \BibitemShut  [1]{\csname bibitem#1\endcsname}%
\let\auto@bib@innerbib\@empty
%</preamble>
\bibitem [{\citenamefont {Onsager}(1931)}]{Onsager1931}%
  \BibitemOpen
  \bibfield  {author} {\bibinfo {author} {\bibfnamefont {L.}~\bibnamefont
  {Onsager}},\ }\href {\doibase 10.1103/PhysRev.37.405} {\bibfield  {journal}
  {\bibinfo  {journal} {Phys. Rev.}\ }\textbf {\bibinfo {volume} {37}},\
  \bibinfo {pages} {405} (\bibinfo {year} {1931})}\BibitemShut {NoStop}%
\bibitem [{\citenamefont {Schnakenberg}(1976)}]{Schnakenberg1976}%
  \BibitemOpen
  \bibfield  {author} {\bibinfo {author} {\bibfnamefont {J.}~\bibnamefont
  {Schnakenberg}},\ }\href {\doibase 10.1103/RevModPhys.48.571} {\bibfield
  {journal} {\bibinfo  {journal} {Rev. Mod. Phys.}\ }\textbf {\bibinfo {volume}
  {48}},\ \bibinfo {pages} {571} (\bibinfo {year} {1976})}\BibitemShut
  {NoStop}%
\bibitem [{\citenamefont {Hill}(1989)}]{Book_Hill1989}%
  \BibitemOpen
  \bibfield  {author} {\bibinfo {author} {\bibfnamefont {T.~L.}\ \bibnamefont
  {Hill}},\ }\href {\doibase 10.1007/978-1-4612-3558-3} {\emph {\bibinfo
  {title} {Free Energy Transduction and Biochemical Cycle Kinetics}}}\
  (\bibinfo  {publisher} {Springer-Verlag New York, Inc.},\ \bibinfo {year}
  {1989})\BibitemShut {NoStop}%
\bibitem [{\citenamefont {Peusner}\ \emph {et~al.}(1985)\citenamefont
  {Peusner}, \citenamefont {Mikulecky}, \citenamefont {Bunow},\ and\
  \citenamefont {Caplan}}]{Peusner1985}%
  \BibitemOpen
  \bibfield  {author} {\bibinfo {author} {\bibfnamefont {L.}~\bibnamefont
  {Peusner}}, \bibinfo {author} {\bibfnamefont {D.~C.}\ \bibnamefont
  {Mikulecky}}, \bibinfo {author} {\bibfnamefont {B.}~\bibnamefont {Bunow}}, \
  and\ \bibinfo {author} {\bibfnamefont {S.~R.}\ \bibnamefont {Caplan}},\
  }\href {\doibase 10.1063/1.449678} {\bibfield  {journal} {\bibinfo  {journal}
  {The Journal of Chemical Physics}\ }\textbf {\bibinfo {volume} {83}},\
  \bibinfo {pages} {5559} (\bibinfo {year} {1985})},\ \Eprint
  {http://arxiv.org/abs/https://pubs.aip.org/aip/jcp/article-pdf/83/11/5559/18956872/5559\_1\_online.pdf}
  {https://pubs.aip.org/aip/jcp/article-pdf/83/11/5559/18956872/5559\_1\_online.pdf}
  \BibitemShut {NoStop}%
\bibitem [{\citenamefont {den Broeck}\ and\ \citenamefont
  {Esposito}(2015)}]{Broeck2015}%
  \BibitemOpen
  \bibfield  {author} {\bibinfo {author} {\bibfnamefont {C.~V.}\ \bibnamefont
  {den Broeck}}\ and\ \bibinfo {author} {\bibfnamefont {M.}~\bibnamefont
  {Esposito}},\ }\href {\doibase 10.1016/j.physa.2014.04.035} {\bibfield
  {journal} {\bibinfo  {journal} {Physica A: Statistical Mechanics and its
  Applications}\ }\textbf {\bibinfo {volume} {418}},\ \bibinfo {pages} {6}
  (\bibinfo {year} {2015})}\BibitemShut {NoStop}%
\bibitem [{\citenamefont {Callen}(1948)}]{Callen1948vol73}%
  \BibitemOpen
  \bibfield  {author} {\bibinfo {author} {\bibfnamefont {H.~B.}\ \bibnamefont
  {Callen}},\ }\href {\doibase 10.1103/physrev.73.1349} {\bibfield  {journal}
  {\bibinfo  {journal} {Physical Review}\ }\textbf {\bibinfo {volume} {73}},\
  \bibinfo {pages} {1349} (\bibinfo {year} {1948})}\BibitemShut {NoStop}%
\bibitem [{\citenamefont {Ioffe}\ \emph {et~al.}(1959)\citenamefont {Ioffe},
  \citenamefont {Stil’bans}, \citenamefont {Iordanishvili}, \citenamefont
  {Stavitskaya}, \citenamefont {Gelbtuch},\ and\ \citenamefont
  {Vineyard}}]{Ioffe1959}%
  \BibitemOpen
  \bibfield  {author} {\bibinfo {author} {\bibfnamefont {A.~F.}\ \bibnamefont
  {Ioffe}}, \bibinfo {author} {\bibfnamefont {L.~S.}\ \bibnamefont
  {Stil’bans}}, \bibinfo {author} {\bibfnamefont {E.~K.}\ \bibnamefont
  {Iordanishvili}}, \bibinfo {author} {\bibfnamefont {T.~S.}\ \bibnamefont
  {Stavitskaya}}, \bibinfo {author} {\bibfnamefont {A.}~\bibnamefont
  {Gelbtuch}}, \ and\ \bibinfo {author} {\bibfnamefont {G.}~\bibnamefont
  {Vineyard}},\ }\href {\doibase 10.1063/1.3060810} {\bibfield  {journal}
  {\bibinfo  {journal} {Physics Today}\ }\textbf {\bibinfo {volume} {12}},\
  \bibinfo {pages} {42} (\bibinfo {year} {1959})}\BibitemShut {NoStop}%
\bibitem [{\citenamefont {Esposito}\ \emph {et~al.}(2015)\citenamefont
  {Esposito}, \citenamefont {Ochoa},\ and\ \citenamefont
  {Galperin}}]{Esposito2015_vol91}%
  \BibitemOpen
  \bibfield  {author} {\bibinfo {author} {\bibfnamefont {M.}~\bibnamefont
  {Esposito}}, \bibinfo {author} {\bibfnamefont {M.~A.}\ \bibnamefont {Ochoa}},
  \ and\ \bibinfo {author} {\bibfnamefont {M.}~\bibnamefont {Galperin}},\
  }\href {\doibase 10.1103/PhysRevB.91.115417} {\bibfield  {journal} {\bibinfo
  {journal} {Phys. Rev. B}\ }\textbf {\bibinfo {volume} {91}},\ \bibinfo
  {pages} {115417} (\bibinfo {year} {2015})}\BibitemShut {NoStop}%
\bibitem [{\citenamefont {Rax}(2015)}]{Book_Rax2015}%
  \BibitemOpen
  \bibfield  {author} {\bibinfo {author} {\bibfnamefont {J.-M.}\ \bibnamefont
  {Rax}},\ }\href@noop {} {\emph {\bibinfo {title} {Physique de la conversion
  d'{\'e}nergie}}}\ (\bibinfo  {publisher} {EDP Science et Edition CNRS},\
  \bibinfo {year} {2015})\BibitemShut {NoStop}%
\bibitem [{\citenamefont {Vroylandt}\ \emph {et~al.}(2018)\citenamefont
  {Vroylandt}, \citenamefont {Lacoste},\ and\ \citenamefont
  {Verley}}]{Vroylandt2018vol2018}%
  \BibitemOpen
  \bibfield  {author} {\bibinfo {author} {\bibfnamefont {H.}~\bibnamefont
  {Vroylandt}}, \bibinfo {author} {\bibfnamefont {D.}~\bibnamefont {Lacoste}},
  \ and\ \bibinfo {author} {\bibfnamefont {G.}~\bibnamefont {Verley}},\ }\href
  {\doibase 10.1088/1742-5468/aaa8fe} {\bibfield  {journal} {\bibinfo
  {journal}
  {\href{https://pperso.ijclab.in2p3.fr/page_perso/Verley/Papers/Vroylandt2018_vol2018.pdf}{J.
  Stat. Mech: Theory Exp.}}\ } (\bibinfo {year} {2018}),\
  10.1088/1742-5468/aaa8fe}\BibitemShut {NoStop}%
\bibitem [{\citenamefont {Raux}\ \emph {et~al.}(2024)\citenamefont {Raux},
  \citenamefont {Goupil},\ and\ \citenamefont {Verley}}]{Raux2024vol110}%
  \BibitemOpen
  \bibfield  {author} {\bibinfo {author} {\bibfnamefont {P.}~\bibnamefont
  {Raux}}, \bibinfo {author} {\bibfnamefont {C.}~\bibnamefont {Goupil}}, \ and\
  \bibinfo {author} {\bibfnamefont {G.}~\bibnamefont {Verley}},\ }\href
  {\doibase 10.1103/PhysRevE.110.014134} {\bibfield  {journal} {\bibinfo
  {journal}
  {\href{https://pperso.ijclab.in2p3.fr/page_perso/Verley/Papers/Raux2024vol110.pdf}{Phys.
  Rev. E}}\ }\textbf {\bibinfo {volume} {110}},\ \bibinfo {pages} {014134}
  (\bibinfo {year} {2024})},\ \Eprint {http://arxiv.org/abs/2309.12922}
  {2309.12922} \BibitemShut {NoStop}%
\bibitem [{\citenamefont {Polettini}\ \emph {et~al.}(2016)\citenamefont
  {Polettini}, \citenamefont {Bulnes-Cuetara},\ and\ \citenamefont
  {Esposito}}]{Polettini2016_vol94}%
  \BibitemOpen
  \bibfield  {author} {\bibinfo {author} {\bibfnamefont {M.}~\bibnamefont
  {Polettini}}, \bibinfo {author} {\bibfnamefont {G.}~\bibnamefont
  {Bulnes-Cuetara}}, \ and\ \bibinfo {author} {\bibfnamefont {M.}~\bibnamefont
  {Esposito}},\ }\href {\doibase 10.1103/PhysRevE.94.052117} {\bibfield
  {journal} {\bibinfo  {journal} {Phys. Rev. E}\ }\textbf {\bibinfo {volume}
  {94}},\ \bibinfo {pages} {052117} (\bibinfo {year} {2016})}\BibitemShut
  {NoStop}%
\bibitem [{\citenamefont {Apertet}\ \emph {et~al.}(2012)\citenamefont
  {Apertet}, \citenamefont {Ouerdane}, \citenamefont {Glavatskaya},
  \citenamefont {Goupil},\ and\ \citenamefont {Lecoeur}}]{Apertet2012}%
  \BibitemOpen
  \bibfield  {author} {\bibinfo {author} {\bibfnamefont {Y.}~\bibnamefont
  {Apertet}}, \bibinfo {author} {\bibfnamefont {H.}~\bibnamefont {Ouerdane}},
  \bibinfo {author} {\bibfnamefont {O.}~\bibnamefont {Glavatskaya}}, \bibinfo
  {author} {\bibfnamefont {C.}~\bibnamefont {Goupil}}, \ and\ \bibinfo {author}
  {\bibfnamefont {P.}~\bibnamefont {Lecoeur}},\ }\href {\doibase
  10.1209/0295-5075/97/28001} {\bibfield  {journal} {\bibinfo  {journal} {EPL
  (Europhysics Letters)}\ }\textbf {\bibinfo {volume} {97}},\ \bibinfo {pages}
  {28001} (\bibinfo {year} {2012})}\BibitemShut {NoStop}%
\bibitem [{\citenamefont {Goupil}\ \emph {et~al.}(2011)\citenamefont {Goupil},
  \citenamefont {Seifert}, \citenamefont {Zabrocki}, \citenamefont
  {M{\"u}ller},\ and\ \citenamefont {Snyder}}]{Goupil2011_vol13}%
  \BibitemOpen
  \bibfield  {author} {\bibinfo {author} {\bibfnamefont {C.}~\bibnamefont
  {Goupil}}, \bibinfo {author} {\bibfnamefont {W.}~\bibnamefont {Seifert}},
  \bibinfo {author} {\bibfnamefont {K.}~\bibnamefont {Zabrocki}}, \bibinfo
  {author} {\bibfnamefont {E.}~\bibnamefont {M{\"u}ller}}, \ and\ \bibinfo
  {author} {\bibfnamefont {G.~J.}\ \bibnamefont {Snyder}},\ }\href {\doibase
  10.3390/e13081481} {\bibfield  {journal} {\bibinfo  {journal} {Entropy}\
  }\textbf {\bibinfo {volume} {13}},\ \bibinfo {pages} {1481} (\bibinfo {year}
  {2011})}\BibitemShut {NoStop}%
\bibitem [{\citenamefont {Kedem}\ and\ \citenamefont
  {Caplan}(1965)}]{Kedem1965_vol61}%
  \BibitemOpen
  \bibfield  {author} {\bibinfo {author} {\bibfnamefont {O.}~\bibnamefont
  {Kedem}}\ and\ \bibinfo {author} {\bibfnamefont {R.~S.}\ \bibnamefont
  {Caplan}},\ }\href {\doibase 10.1039/TF9656101897} {\bibfield  {journal}
  {\bibinfo  {journal} {Trans. Faraday Soc.}\ }\textbf {\bibinfo {volume}
  {61}},\ \bibinfo {pages} {1897} (\bibinfo {year} {1965})}\BibitemShut
  {NoStop}%
\bibitem [{\citenamefont {Polettini}\ \emph {et~al.}(2015)\citenamefont
  {Polettini}, \citenamefont {Verley},\ and\ \citenamefont
  {Esposito}}]{Polettini2015_vol114}%
  \BibitemOpen
  \bibfield  {author} {\bibinfo {author} {\bibfnamefont {M.}~\bibnamefont
  {Polettini}}, \bibinfo {author} {\bibfnamefont {G.}~\bibnamefont {Verley}}, \
  and\ \bibinfo {author} {\bibfnamefont {M.}~\bibnamefont {Esposito}},\ }\href
  {\doibase 10.1103/PhysRevLett.114.050601} {\bibfield  {journal} {\bibinfo
  {journal}
  {\href{https://pperso.ijclab.in2p3.fr/page_perso/Verley/Papers/Polettini2015_vol114.pdf}{Phys.
  Rev. Lett.}}\ }\textbf {\bibinfo {volume} {114}},\ \bibinfo {pages} {050601}
  (\bibinfo {year} {2015})}\BibitemShut {NoStop}%
\bibitem [{\citenamefont {Apertet}\ \emph {et~al.}(2013)\citenamefont
  {Apertet}, \citenamefont {Ouerdane}, \citenamefont {Goupil},\ and\
  \citenamefont {Lecoeur}}]{Apertet2013_vol88}%
  \BibitemOpen
  \bibfield  {author} {\bibinfo {author} {\bibfnamefont {Y.}~\bibnamefont
  {Apertet}}, \bibinfo {author} {\bibfnamefont {H.}~\bibnamefont {Ouerdane}},
  \bibinfo {author} {\bibfnamefont {C.}~\bibnamefont {Goupil}}, \ and\ \bibinfo
  {author} {\bibfnamefont {P.}~\bibnamefont {Lecoeur}},\ }\href {\doibase
  10.1103/PhysRevE.88.022137} {\bibfield  {journal} {\bibinfo  {journal} {Phys.
  Rev. E}\ }\textbf {\bibinfo {volume} {88}},\ \bibinfo {pages} {022137}
  (\bibinfo {year} {2013})}\BibitemShut {NoStop}%
\end{thebibliography}%

\end{document}